\documentstyle[12pt,epsf]{article}
\newcommand{\rf}[1]{(\ref{#1})}
\newcommand{\beq}{\begin{equation}}
\newcommand{\eeq}{\end{equation}}
\newcommand{\bea}{\begin{eqnarray}}
\newcommand{\eea}{\end{eqnarray}}
\newcommand{\nn}{\nonumber \\}

\newcommand{\G}{{\rm G}}

\newcommand{\gut}{\alpha_{\rm GUT}}

\def\gsim{\raise.3ex\hbox{$>$\kern-.75em\lower1ex\hbox{$\sim$}}}
\def\lsim{\raise.3ex\hbox{$<$\kern-.75em\lower1ex\hbox{$\sim$}}}
\def\sepand{\rule{10cm}{0pt}\and}

\begin{document}
\topmargin 0cm
\topskip 0mm
 {\title
{\null\vskip-3truecm
{ \hskip10truecm {\small NORDITA-94/6 P\hfill }\vskip 1cm}
{\bf Ferromagnetic vacuum and galactic magnetic fields}}
\author{
{\sc Kari Enqvist$^1$} \\
{\sl Nordita} \\
{\sl Blegdamsvej 17,
DK-2100 Copenhagen, Denmark}\\
and\\
{\sc Poul Olesen$^2$ }\\
{\sl The Niels Bohr Institute, University of Copenhagen} \\
{\sl Blegdamsvej 17, DK-2100 Copenhagen, Denmark} \\
\sepand
}\maketitle}

\begin{abstract}
\noindent
Non-abelian gauge theories may have a ferromagnet-like vacuum with a
non-zero magnetic field, which also exists at finite temperature.
We argue that the formation of the ferromagnet-like vacuum at GUT
scales gives rise to a Maxwell magnetic field imprinted on
the comoving plasma, and that it is energetically favourable to
align the field
in different correlation volumes. This results in a coherent magnetic
field pertaining the whole universe, with a magnitude
   $B_{now}\simeq 10^{-14}\G$, which is of the correct size to serve as
the seed field for the galactic dynamo.
\end{abstract}
\vfil
\footnoterule
{\small  $^1$enqvist@nbivax.nbi.dk;  $^2$polesen@nbivax.nbi.dk}
\thispagestyle{empty}
\newpage
\setcounter{page}{1}

The nearby galaxies have magnetic fields of the order
of $10^{-6}$ G \cite{dynamo},
and a field of a similar magnitude has recently been observed also in a
object with z=0.395 \cite{bigz}. Such a magnetic
field can be understood as an
amplification of a  weak seed field of the order of $10^{-18}$ G
(with an uncertainty of a few orders of magnitude) on a comoving
scale 100 kpc . One interesting possibility is that the seed field
is truly primordial, with an origin that predates nucleosynthesis.
 This would then have important
consequences for Dirac neutrino properties
\cite{viki}, as well as
for the primordial nucleosynthesis, which constrains \cite{cheng}
the magnetic field to be less than about $10^{11}$ G at $t\simeq 1$ min and
at the scale $10^4$ cm.

{}From a theoretical point of view, the generation of a sufficiently
large persistent
magnetic field in the early universe is rather difficult. There are
various attempts, relying on phase transitions such as the cosmic inflation
\cite{attempts}, but the field often  comes out to be too small to be of
cosmological interest. It has however been suggested that a large
field might actually be generated at the electroweak phase transition
because of  random fluctuations in the
 Higgs field gradients \cite{vachaspati}.
If one assumes a stochastic, uncorrelated distribution of the gradients,
or of the magnetic field itself, one finds \cite{eo} today
at 100 kpc a root--mean--square field of the order of $10^{-19}$ G,
which could well be the seed field. This positive result is based
on calculating the statistical averages along an arbitrary curve.
This is not the only possibility, but averaging over areas or volumes
would produce a field far too small to be of relevance for the dynamo
effect. It is therefore of interest also to look for other mechanisms.

In the present paper we consider another possibility, which is based on
the observation that, due to quantum fluctuations,
the Yang--Mills vacuum is unstable in a large enough background magnetic
field \cite{savvidy}. We propose that magnetic field fluctuations in
the early universe are sufficient to trigger the phase transition to
a new, ferromagnet--like ground state with a magnetic field made
permanent by charged plasma.

At zero temperature the one--loop effective energy
for constant background magnetic fields in the pure SU(N) theory has
in general a non--trivial
vacuum, the Savvidy vacuum, corresponding to a non--zero magnetic field
\cite{savvidy}. This is an analog of a ferromagnet. To phrase the
result in a general manner we may use the language of the renormalization
group. The effective Lagrangian is then computed as a function of
${\rm Tr}\; F^2$ in the case where ${\rm Tr}\; F^2$ is
space--time independent. The result obtained in the Savvidy case
corresponds to the assumption that the $\beta$--function has a
Landau singularity, i.e. one assumes that
\beq
\left\vert~\int^{\infty}_g{dx\over\beta (x)}~\right\vert <\infty.
\eeq
Then the effective Lagrangian has a minimum away from the perturbative
ground state ${\rm Tr}\; F^2=0$, given by
\beq
\frac 12g^2{\rm Tr}\; F^2_{\mu\nu}\vert_{\rm min}=\Lambda^4,
\eeq
where $\Lambda$ is the renormalization group invariant scale
\beq
\Lambda=\mu\exp\left(-\int_\infty^g{dx\over\beta (x)}\right).
\eeq
The condition for the minimum can be realized in many ways. One is to
take a constant non--abelian magnetic field $B^a_i=\epsilon_{ijk}F^a_{jk}$
with a non--zero component only in one direction in the group space,
and with a length given by
\beq
g\sqrt{B^aB^a}=\Lambda^2.
\eeq
Another realization would be to assume that
$A^a_j=C\delta^a_j$. Such a field is,
however, not covariantly constant but $D_\mu^{ab}F^b_{i\mu}=2g^2C^3\delta^a_i$,
and hence it is not a suitable background field for the effective Lagrangian
method.

Consider now SU(N) at the one--loop level. We then have
the one--loop, zero temperature  effective energy for a constant
background non--abelian magnetic field which in pure SU(N) theory reads
\cite{savvidy}
\beq
V(B)=\frac 12 B^2+ \frac{11N}{96\pi^2}g^2B^2\left(\ln {gB\over\mu^2}-
\frac 12\right)              \label{1}
\eeq
with a minimum at
\beq
gB_{\rm min}=\mu^2\exp \left(-{48\pi^2\over 11Ng^2}\right)  \label{2}
\eeq
and $V_{\rm min}\equiv V(B_{\rm min})=-0.029(gB_{\rm min})^2$.
Thus the ground state (the Savvidy vacuum)
has a non--zero  non--abelian magnetic field, the magnitude
of which is exponentially suppressed relative to the renormalization
scale, or the typical momentum scale of the system. Thus, for example,
for SU(2)$_{\rm L}$ at the electroweak scale the vacuum magnetic field
would be very small.
In the early
universe, however, where possibly a grand unified symmetry is
valid, the exponential suppression is less severe. It is also
attenuated by the running of the coupling constant. For a set of
representative
numbers, one might consider an SU(5) model with $\gut\simeq 1/25$
and $T_{\rm GUT}\simeq 10^{15}$ GeV, as in
the supersymmetric Standard Model. This yields $B\simeq 5\times 10^{-8}\mu^2$,
which, as we shall argue, is potentially important.

In the early universe the effective energy picks up thermal corrections
from fermion\-ic, gauge boson, and Higgs boson loops. In SU(2) these are
obtained by summing the Boltzmann factors $\exp(-\beta E_n)$ for the
oscillator modes
\beq
E_n^2=p^2+2gB(n+\frac 12)+2gBS_3+m^2(T), \label{21}
\eeq
where $S_3=\pm 1/2~(\pm 1)$ for fermions (vectors bosons).
In Eq. \rf{21} we have included the thermally induced mass $m(T)\sim gT$,
 corresponding to
a ring summation of the relevant diagrams. Numerically, the effect of
the thermal mass turns out to be very important.

The detailed form of
the thermal correction depends on the actual model, but we may take our
cue from the SU(2) one--loop calculation, which for the
fermionic and scalar cases can be extracted from
the real--time QED calculation in \cite{thermal}. The result is
\bea
\delta V_T^f&=&\frac
{(gB)^2}{4\pi^2}\sum_{l=1}^{\infty}(-1)^{l+1}\int_0^{\infty}
\frac{dx}{x^3}e^{-K_l^a(x)}\left[\; x{\rm coth}(x)
-1\right], \nn
\delta V_T^s&=&\frac {(gB)^2}{8\pi^2}\sum_{l=1}^{\infty}\int_0^{\infty}
\frac{dx}{x^3}e^{-K_l^a(x)}\left[\; x{1\over
{\rm sinh}(x)}-1\right],
\label{3}
\eea
where the normalization is such that
the correction vanishes for zero field, and
\beq
K_l^a(x)={gBl^2\over 4xT^2}+{m_a^2x\over gB}
\eeq
where $a=f,~b$ stands for fermions or bosons.

For vector bosons there is the added complication that there exists a
negative, unstable mode, which gives rise to an imaginary part\footnote{
Physically the  imaginary part is an indicator that the
vacuum also contains vector particles \cite{no}.}.
At high temperatures the instability is absent for fields such that
$gB<m^2(T)$, which is the case we are interested in here, so that no
regulation of the unstable $n=0,~S_3=-1$ mode is needed.
Thus  we find
\beq
\delta V_T^v=\frac {(gB)^2}{8\pi^2}\sum_{l=1}^{\infty}\int_0^{\infty}
\frac{dx}{x^3} e^{-K_l^b(x)}\left[\;
x{{\rm cosh}(2x)\over {\rm sinh}(x)}-1\right].  \label{4}
\eeq


At high temperature, the bosonic contributions are more important than
the fermionic ones. When $B\ll T^2\simeq m^2(T)$, we find numerically that
 $\delta V_T^v\simeq 0.016\times (gB)^2$.
This gives rise to a small correction to the magnitude of the field at
the minimum as obtained from Eq. \rf{2}. It only alters the
classical quadratic term in Eq. \rf{1}, and replacing $B^2/2$ by $cB^2$
one finds that the minimum  occurs
at $gB_{\rm min}=\mu^2\exp (-96c\pi^2/11Ng^2)$.
Thus in our case the vector boson correction would change the magnitude
of the field at the minimum by 10\%, which is a negligible amount. In
these considerations we have set $\mu\simeq T$
since this is the typical scale of the
particle momenta in the thermal bath. We may thus conclude that the
Savvidy vacuum  exists for all $T$.

The transition
to this new ferromagnet-like vacuum is triggered by local fluctuations.
Such local fluctuations of the magnetic field
are in fact very likely to occur due to currents generated in the
primeval plasma. To see this, consider e.g.  quarks traversing
the plasma and generating a current ${\bf j}=\nabla \times {\bf B}$.
The typical interparticle distance is $L\sim 1/T$ and a typical curl
goes like $1/L$
 so that $B\sim jL$. The quarks move with the speed of light, so that
$j$ is like charge density, and  there is one charge in the  volume
$L^3$. Thus the Maxwell equations imply that $B\sim 1/L^2=T^2$.

Such a local field will normally be dissipated very rapidly
due to collisions
between the particles in the plasma. Note, however, that in the very
early universe the particle reactions become slower and slower as
compared with the expansion rate of the universe; for example, it
has been calculated that in QCD quarks no longer are in full thermal
contact when $T\simeq 3\times 10^{14}$ GeV \cite{sirkka}. Therefore at
GUT temperatures we expect to have persistent currents and magnetic
fields of the order of $T^2$ over length (time) scales almost comparable
to the horizon size (age of the universe).

Thus a local magnetic bubble will trigger the creation of the
Savvidy vacuum inside a given particle horizon at scales $\mu \simeq T$.
A constant  non--abelian magnetic field, given by \rf{2},
is then imprinted on the
plasma of particles carrying the relevant charges.
The flux remains
conserved (the primodial plasma is an extremely good conductor), and we may
write
\beq
B(T)=g_{\rm GUT}^{-1}\mu^2\exp \left(-{48\pi^2\over 11Ng^2}\right)
\left({T^2\over \mu^2}\right)\simeq
3\times 10^{42}\G\left({R(t_{\rm GUT})\over R(t)}\right)^2, \label{5}
\eeq
where $R(t)$ is the scale factor of the universe, and the last figure is
for susy SU(5). The expression \rf{5}
is valid in a horizon
volume with an initial size $l_0\simeq 10^{-28}$ m. The  Maxwell
magnetic field $B_{em}$ is a projection in the space of non--abelian
magnetic
fields, and we take it to be of the  size comparable to $B$ in Eq. \rf{5}.
Note that the magnetic energy per horizon is much less than the radiation
energy.

As time passes, the
universe undergoes a number of phase transitions. Each of these correspond
to new types of ferromagnetic vacua,
which in general have decreasing field strengths.
However, the original GUT vacuum has existed for a time that
is long enough for the plasma to interact with the vacuum field $B$
given by \rf{5}. This interaction does not allow for the GUT flux to decrease
once it has been created since it has become a feature of the plasma, which
conserves the flux in the sense that the magnetic lines of force are
frozen into the fluid. Eq. \rf{5} expresses this in comoving coordinates.
Thus we expect that even today the GUT field should be present, at
least in the regions which contain plasma.

{}From Eq. \rf{5} we find that the Maxwell magnetic field at
$t_{now}\simeq 10^{10}$ yr is given
by $B_{now}\simeq 3\times 10^{42}\G (t_{\rm GUT}/t_*)(t_*/t_{now})^{4/3}
\simeq 10^{-14}\G$. Here we have assumed that the change to matter
dominated universe  takes place at $t_*\simeq 8750$ yr, corresponding
to $\Omega h^2\simeq 0.4$. Such a magnetic field appears comparable
to what is needed for the seed field in galactic dynamo models.
Note also that at nucleosynthesis one obtains $B\sim 10^4$ G, which is well
below the nucleosynthesis bound on magnetic fields \cite{cheng}.

However, the GUT causal domain $l_0$ has  today the size of only about
1 m.
Obviously, during the course of the evolution of the universe, domains
with magnetic fields pointing to different directions have come into
contact with each other. This results in domain walls between
the magnetic bubbles where extra energy is stored. Since
the energy inside the magnetic bubble is independent of the orientation
of the field, it
is energetically favourable to radiate away  the walls and align  the
fields in
different domains.

At first this might seem peculiar considering that in ordinary ferromagnetic
materials, in the absence of strong external fields, a domain
structure is actually favoured. This is, however, connected to the
fact that real ferromagnets are finite in size. Suppose we have a
finite ferromagnet which is magnetized in one direction. The resulting
magnetic field external to the ferromagnet then has a considerable
energy. This situation compares unfavourably to the case where the
domains are present: by suitably arranging the domains, the field external
to the magnet can be completely avoided, at the expense of surface
energy stored in the domain walls. Since the "external energy" is
a volume effect, the price paid for the walls (which are surface effects)
is smaller than the gain obtained by avoiding the "external energy".

In our case there is nothing external to the ferromagnet--like
 vacuum, which fills the whole universe. Hence we conclude that
it is energetically favourable to remove the domain walls. This results
in a homogeneous magnetic field pertaining the whole universe.
The size of the field is determined by the scale at which the
ferromagnet vacuum is created, and the earlier this happens, the bigger
the field. If there is a period of cosmic inflation, then the
relevant field would be created after reheating. If the
reheating temperature is comparable to GUT scales, the
strength of the magnetic field would be given as in \rf{5}, with
interesting consequences for the formation of galactic magnetic fields.
In this scenario the primordial seed field is thus a relic from the GUT era.
It would of course be of great interest to observe this relic field
directly in the intergalactic space.
\vskip1truecm\noindent
{\Large\bf Acknowledgements}
\vskip0.5truecm
K.E. wishes to thank Per Elmfors for useful discussions on loop
corrections in the background of a magnetic field.

\end{document}